\documentclass[authoryear, review, 12pt, 3p]{elsarticle}

\usepackage{amssymb}

\usepackage{algorithm2e}
\usepackage{amsmath}

\usepackage{color,soul}

\journal{Engineering Applications of Artificial Intelligence}

\begin{document}

\begin{frontmatter}

\title{A neural attention model for speech command recognition}

\author[label1]{Douglas Coimbra de Andrade}
\address[label1]{Laboratory of Voice, Speech and Singing, Federal University of the State of Rio de Janeiro}
\ead{douglas@cmsoft.com.br}

\author[label2]{Sabato Leo}
\address[label2]{Adecco Italia S.P.A - GSK Vaccines Srl}
\ead{sabato.leo@gmail.com}

\author[label3]{Martin Loesener Da Silva Viana}
\address[label3]{CERN}
\ead{martin.eduard.loesener@cern.ch}

\author[label3]{Christoph Bernkopf}
\ead{christoph.bernkopf@cern.ch}

\begin{abstract}
This paper introduces a convolutional recurrent network with attention for speech command recognition. Attention models are powerful tools to improve performance on natural language, image captioning and speech tasks. The proposed model establishes a new state-of-the-art accuracy of 94.1\% on Google Speech Commands dataset V1 and 94.5\% on V2 (for the 20-commands recognition task), while still keeping a small footprint of only 202K trainable parameters. Results are compared with previous convolutional implementations on 5 different tasks (20 commands recognition (V1 and V2), 12 commands recognition (V1), 35 word recognition (V1) and left-right (V1)). We show detailed performance results and demonstrate that the proposed attention mechanism not only improves performance but also allows inspecting what regions of the audio were taken into consideration by the network when outputting a given category.

\end{abstract}


\end{frontmatter}


\section{Introduction}
\label{secIntro}

Currently, many human-computer interfaces (HCI) like Google Assistant, Microsoft Cortana, Amazon Alexa, Apple Siri and others rely on speech recognition. There is a significant amount of research in the field by all major companies, notably Google and Baidu (\cite{DBLP:journals/corr/AmodeiABCCCCCCD15}, \cite{DBLP:journals/corr/abs-1712-01769}). However, these systems rely on powerful neural network models that usually run in the cloud due to the computation required to transform speech to text, perform natural language processing of user intent and react appropriately. Simple commands, such as ``stop'' and ``go'', that could be processed locally, go through the same processing stages. Therefore, industry applications which do not have the benefit of uninterrupted broadband internet connection cannot incorporate speech recognition in the same manner. The development of lightweight speech command models would enable the development of a multitude of novel engineering applications, such as voice-controlled robots for critical missions and assistive devices that can operate in areas without internet coverage. This aspect is particularly important when designing microcontrollers that support voice-driven commands (\cite{DBLP:journals/corr/abs-1711-07128}).


This work introduces a novel attention-based recurrent network architecture designed to recognize simple speech commands, while still generating a lightweight model that can be loaded in mobile devices and run locally.

The main contributions of this work are:

\begin{enumerate}
    \item Design of a novel recurrent architecture with attention that achieves state-of-the-art performance in command recognition and language identification from speech and is small enough to be run locally;
    \item Visualization of attention weights and discussion about how attention improves accuracy and makes the speech recognition model explainable;
    \item Source code (to be made available at https://github.com/.... after acceptance -- blind review).
\end{enumerate}


Results are presented using Google Speech Command datasets V1 and V2. For complete details about these datasets, refer to \cite{DBLP:journals/corr/abs-1804-03209}.

This paper is structured as follows: Section~\ref{secRelWork} discusses previous work on command recognition and attention models. Section~\ref{secNNModels} presents the proposed neural network architecture. Section~\ref{secRes} shows results obtained on various tasks related to Google Speech Command datasets V1 and V2, as well as attention and confusion matrix plots. Section~\ref{secConclusion} summarizes this work and presents possible directions for future work.


\subsection{Related Work}
\label{secRelWork}

Identification of speech commands, also known as keyword spotting (KWS), is important from an engineering perspective for a wide range of applications, from indexing audio databases and indexing keywords (\cite{TABIBIAN20131660}, \cite{SANGEETHA2014287}) to running speech models locally in microcontrollers (\cite{DBLP:journals/corr/abs-1711-07128}).

The development of neural attention models (\cite{DBLP:journals/corr/BahdanauCB14}, \cite{DBLP:journals/corr/VaswaniSPUJGKP17}) increased performance on multiple tasks, especially those related to long sequence to sequence models. These models are extremely powerful ways to understand what parts of the input are being used by the neural network to predict outputs, as shown in the case of image captioning (\cite{DBLP:journals/corr/XuBKCCSZB15}). In the case of acoustic models, Connectionist Temporal Classification (CTC) loss shows good performance in English and Mandarin speech to text tasks (\cite{DBLP:journals/corr/AmodeiABCCCCCCD15}). There is also work on sequence discriminative frameworks (\cite{CHEN2018100}). However, to the best of our knowledge, attention for single word recognition has not been investigated.

Command recognition using deep residual networks has been investigated in \cite{DBLP:journals/corr/abs-1710-10361}, \cite{DBLP:journals/corr/ArikKCHGFPC17} and \cite{Sainath2015ConvolutionalNN}. In particular, the problem of limited size architectures has been extensively explored in \cite{DBLP:journals/corr/abs-1711-07128}. The best results achieve over 95\% accuracy in specific tasks using spectrogram methods. However, most models perform significantly worse than the proposed recurrent neural network (RNN) with attention even with a greater number of parameters. Moreover, the proposed attention mechanism makes results explainable and easy to interpret, which is fundamental for engineering applications and partly solves the problem of ``black box in deep learning'' (\cite{DBLP:journals/corr/abs-1805-08355}). Results using raw waveform without any Fourier analysis have also been investigated (\cite{Jansson2018}).

Results are presented on accuracy of left vs right (i.e., identifying only the words ``left'', ``right'' or none of those two), 20 commands and 10 non-commands and all 35 words (\cite{DBLP:journals/corr/abs-1710-08377}). Note that while the first version of the dataset only has 30 words in the test set, V2 has all 35 (denoted by  ``35-word task'' in this paper).




\section{Neural Network Implementation}
\label{secNNModels}

The Keras interface (\cite{chollet2015keras}) was used to implement all neural networks on top of Tensorflow backend (\cite{tensorflow2015-whitepaper}). On top of that, Python library kapre (\cite{choi2017kapre}) is used to provide Keras layers for mel-scale spectrogram computation. The use of kapre library provides extreme versatility to change spectrogram and mel-frequency parameters without having to preprocess the audio in any way. As a result, the inputs to the models are raw WAV files converted to numpy arrays for efficient load with Keras generators.



\subsection{Proposed Architecture}
\label{secPropArchit}

Since the audio files contain a single word command that can be anywhere in the 1$s$ length of the WAV file, it is reasonable to assume that any model that is able to classify an audio should also be able to find what is the appropriate region of interest. Thus, the attention mechanism seems appropriate to this particular task.

The model starts by computing the mel-scale spectrogram of the audio using non-trainable layers implemented in the kapre library. The input to the model is the raw WAV data with original sampling rate of $\sim$16 $kHz$. Mel-scale spectrogram is computed using 80-band mel scale, 1024 discrete Fourier transform points and hop size of 128 points ($\sim$8 $s$). Similar parameters have been successfully used for audio synthesis (\cite{DBLP:journals/corr/WangSSWWJYXCBLA17}) and we noticed that they preserve the visual aspect of the voice formants in the spectrogram (which would allow a human specialist to evaluate the sound - \cite{Sundberg2013278}, \cite{Sundberg2015418}).

After mel-scale spectrogram computation, a set of convolutions is applied to the mel-scale spectrogram (2D output) only in the time dimension to extract local relations in the audio file. A set of two bidirectional long short term memory (LSTM - \cite{Hochreiter:1997:LSM:1246443.1246450}) units is used to capture two-way (forward and backward) long term dependencies in the audio file.

At this point, one of the output vectors of the last LSTM layer is extracted, projected using a dense layer and used as query vector to identify what part of the audio is the most relevant. We choose to use the middle vector of the LSTM output since the voice command is expected to be centered in the audio files. This choice is arbitrary and any vector should work since the double stacked LSTM layers should be able to carry enough ``memory''.

\begin{figure}[!ht]
\centering
\includegraphics[width=1\textwidth]{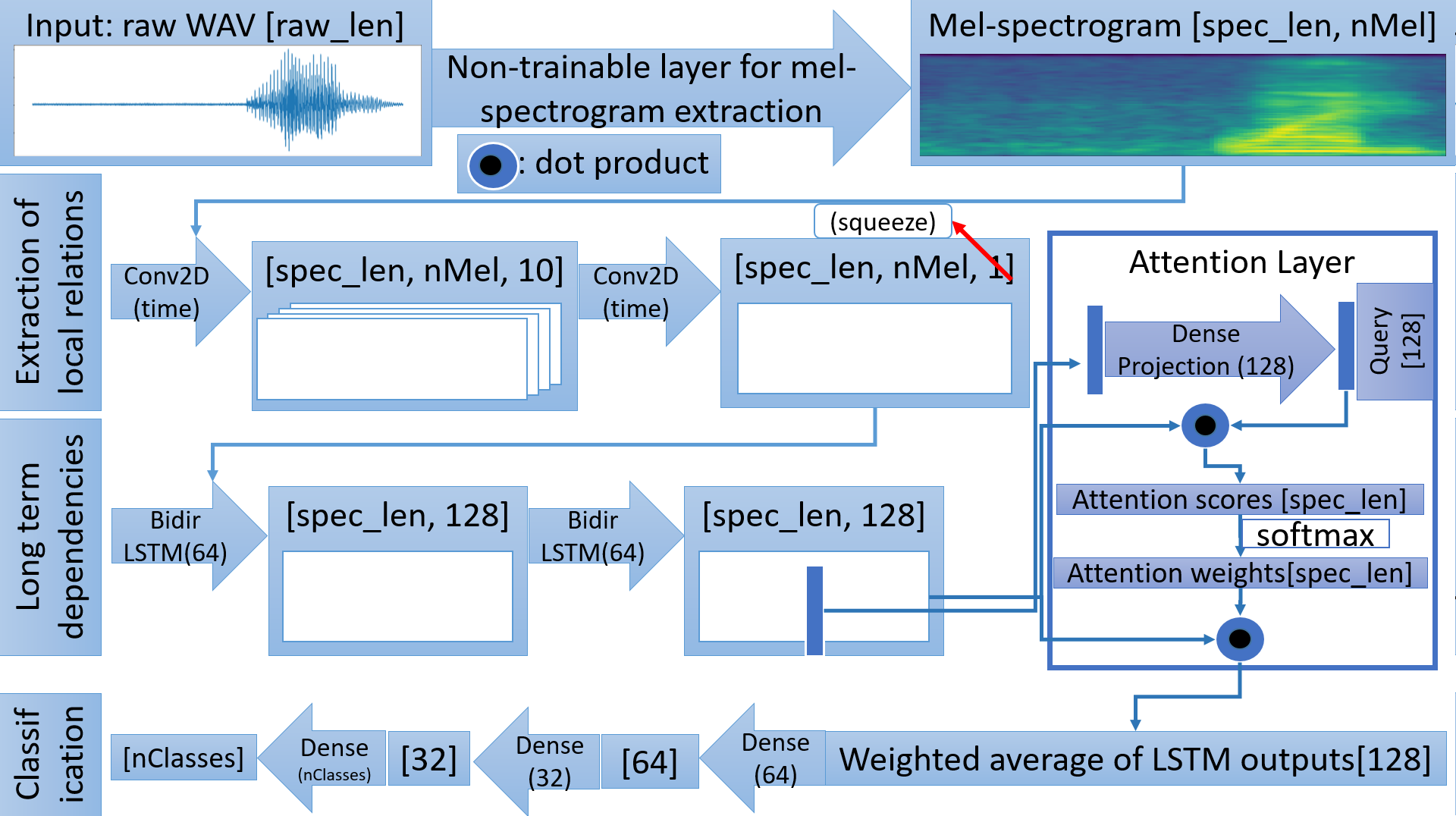}
\caption{Proposed architecture: recurrent neural network with attention mechanism. Numbers between [brackets] are tensor dimensions. raw\_len is WAV audio length (16000 in this case). spec\_len is the sequence length of the generated mel-scale spectrogram. nMel is the number of mel bands. nClasses is the number of desired classes. The activation of the last Dense layer is softmax. The activation of the 64 and 32 dense classification layers is the rectified linear unit (relu).}
\label{figAttRNN}
\end{figure}

Finally, the weighted average of the LSTM output is fed into 3 fully connected layers for classification. Figure~\ref{figAttRNN} summarizes the architecture. For complete details about the implementation, please refer to the repository.

\section{Results}
\label{secRes}

To facilitate comparison with previous results, 5 different tasks were considered:

\begin{itemize}
    \item Recognition of 20 speech commands (+unknown) using Google Speech Dataset V2;
    \item Recognition of 20 speech commands (+unknown) using Google Speech Dataset V1;
    \item Recognition of 12 speech commands (+unknown) using Google Speech Dataset V1;
    \item Recognition of all 35 words (+unknown, only silence samples in this case) using Google Speech Dataset V1;
    \item Recognition of left-right words using Google Speech Dataset V1;
\end{itemize}

For each task, the proposed Attention RNN model was trained for a maximum of 40 epochs. The model with the best accuracy performance on the validation set was saved and training was stopped if no improvement was made in 10 consecutive epochs. Training was done using the ``adam'' algorithm (\cite{DBLP:journals/corr/KingmaB14}) with initial learning rate of 0.001 and decay of 0.4 every 10 epochs. The batch size used was 64. Tests in multiple training runs (3 to 5, depending on the task) show that the training procedure is consistent and standard deviation of accuracy results is $0.2\%$ for all models. Each epoch takes approximately 180~s (V2) and 100~s (V1) to run in a Tesla K80 GPU.

Attention plots and confusion matrices are also presented to allow for better comparison of the results.

\subsection{Speech Command Recognition Accuracy}
\label{secSpeechCmdRecog}

The results obtained with the attention-RNN model are compared with left-right accuracy, 20 command accuracy and 35 word accuracy (\cite{DBLP:journals/corr/abs-1710-08377}), as shown in Table~\ref{tblAccComp}. Attention provides a substantial improvement on these tasks when compared to other models. On the V2 dataset, our results are 94.5\% (20-cmd) and 93.9\% (35-word) -- significantly better than the 20-cmd baseline of 88.2\% from \cite{DBLP:journals/corr/abs-1804-03209}.

\begin{table}[]
\centering
\caption{Accuracy results on the Google Speech Command Dataset V1. DenseNet-101 results from \cite{DBLP:journals/corr/abs-1710-08377}. ConvNet results from \cite{DBLP:journals/corr/abs-1804-03209}. Our attention Model results on the Google Speech Command Dataset V2 are also reported in the last row.}
\label{tblAccComp}
  
\begin{tabular}{l|l|l|l|}
\cline{2-4}
                                                                                                                         & \multicolumn{3}{c|}{\textbf{Accuracy (\%)}}                                                           \\ \hline
\multicolumn{1}{|c|}{\textbf{Model}}                                                                                              & \multicolumn{1}{c|}{20-cmd} & \multicolumn{1}{c|}{35-word} & \multicolumn{1}{c|}{left/right} \\ \hline
\multicolumn{1}{|l|}{DenseNet-121 No pretrain, no multiscale}                                                            & 81.32                       & 80.13                        & 89.19                           \\ \hline
\multicolumn{1}{|l|}{\begin{tabular}[c]{@{}l@{}}DenseNet-121 Pretrained on UrbanSound8K,\\   no multiscale\end{tabular}} & 82.48                       & 81.55                        & 91.40                           \\ \hline
\multicolumn{1}{|l|}{DenseNet-121 No pretrain,  multiscale}                                                              & 82.22                       & 82.11                        & 88.54                           \\ \hline
\multicolumn{1}{|l|}{\begin{tabular}[c]{@{}l@{}}DenseNet-121 Pretrained on UrbanSound8K,\\   multiscale\end{tabular}}    & 85.52                       & 84.35                        & 95.32                           \\ \hline
\multicolumn{1}{|l|}{ConvNet}                                                                                            & 85.4                        & N/A                          & N/A                             \\ \hline
\multicolumn{1}{|l|}{Attention RNN (Ours)}                                                                               & \textbf{94.1}               & \textbf{94.3}                   & \textbf{99.2}                      \\ \hline
\multicolumn{1}{|l|}{Attention RNN (Ours, V2)}                                                                               & \textbf{94.5}               & \textbf{93.9}                   & \textbf{99.4}                      \\ \hline
\end{tabular}
\end{table}

Table~\ref{tblAccComp12} compares results of the attention RNN model when used to recognize only the 12 commands originally proposed in the Kaggle competition (\cite{TFSpRecChallenge}). Results are also presented for attention RNN trained and tested on V2 dataset.

\begin{table}[]
\centering
\caption{Accuracy results on 12-commands from Google Speech Command Dataset V1. ``res'' model results from \cite{DBLP:journals/corr/abs-1804-03209}. ConvNet on raw WAV results from \cite{Jansson2018}.  Depthwise Separable Convolutional Neural Network (DS-CNN) from \cite{DBLP:journals/corr/abs-1711-07128}. }
\label{tblAccComp12}
\begin{tabular}{|l|l|l|}
\hline
\multicolumn{1}{|c|}{\textbf{Model}} & \multicolumn{1}{c|}{\textbf{Accuracy (\%)}} & \multicolumn{1}{c|}{\textbf{Trainable Parameters}} \\ \hline
res15                                & 95.8                                        & 238K                                               \\ \hline
res26                                & 95.2                                        & 438K                                               \\ \hline
res8                                 & 94.1                                        & 110K                                               \\ \hline
ConvNet on raw WAV                                 & 89.4                                        & 700K                                               \\ \hline
DS-CNN                                 & 95.4                                        & 498K                                               \\ \hline
Attention RNN (Ours)                 & \textbf{95.6}                                        & 202K                                               \\ \hline
Attention RNN (Ours, V2)                 & \textbf{96.9}                                        & 202K                                               \\ \hline
\end{tabular}
\end{table}

Furthermore, results obtained using the entire training dataset are compared with leaderboard results from Kaggle competition (\cite{TFSpRecChallenge}) using the held-out data released after the competition ended (\cite{DBLP:journals/corr/abs-1804-03209}). Had this been an entry, the leaderboard score would have been 92.8\%, which would rank \#1 (assuming that the test set released is identical to the one used for scoring).

\subsection{Attention Plots}
\label{secAttPlots}

One of the main advantages of deep learning models with attention is the possibility to explain the results and get intuition about what the model does. Like in the case of images (\cite{DBLP:journals/corr/XuBKCCSZB15}), plotting the attention weights allows visualization of what parts of the audio were most relevant for the classification.

\begin{figure}[!ht]
\centering
\includegraphics[width=1\textwidth]{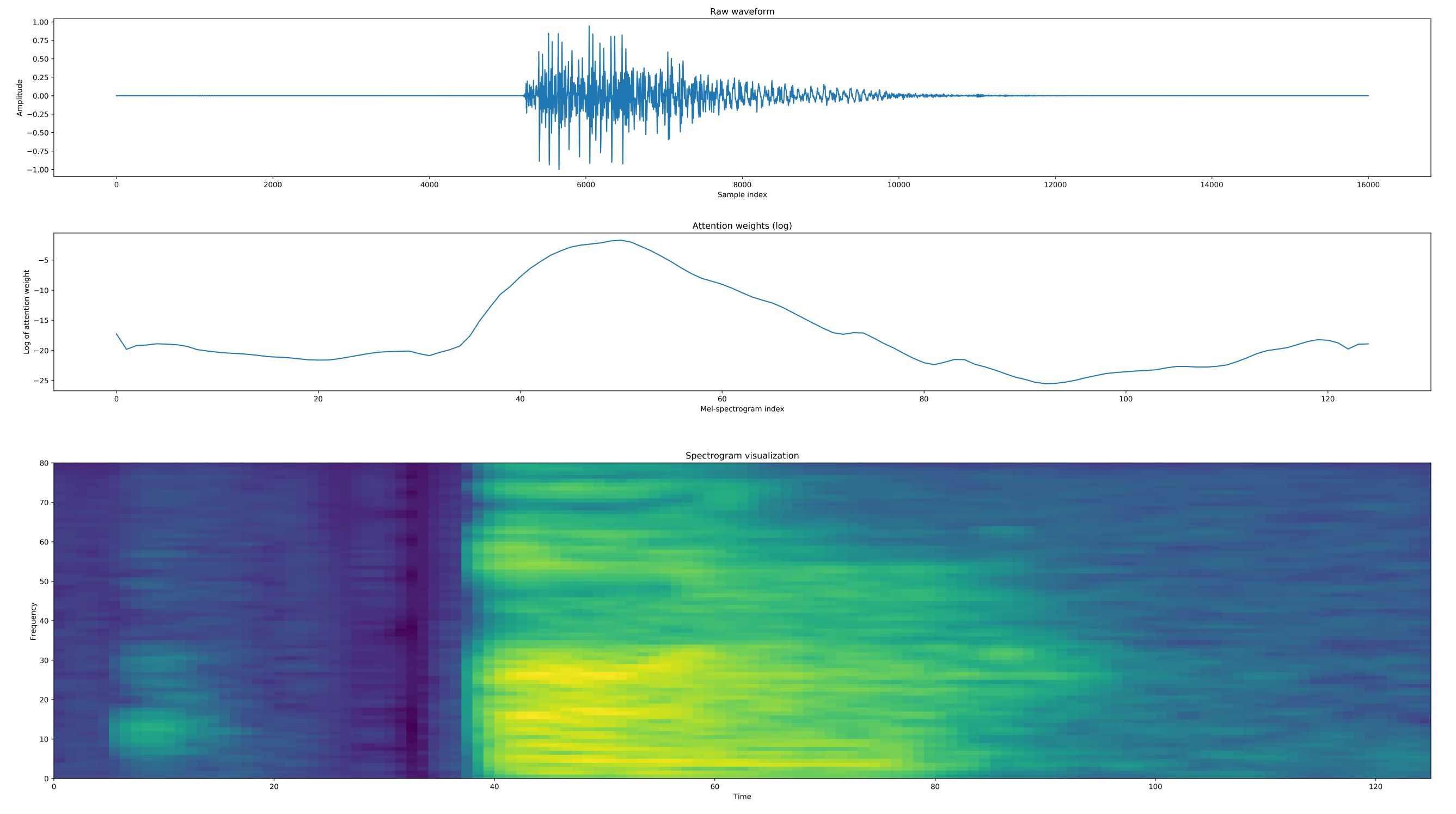}
\caption{Waveform, mel-frequency spectrogram and attention weights for the word ``on''}
\label{figAttON}
\end{figure}

\begin{figure}[!ht]
\centering
\includegraphics[width=1\textwidth]{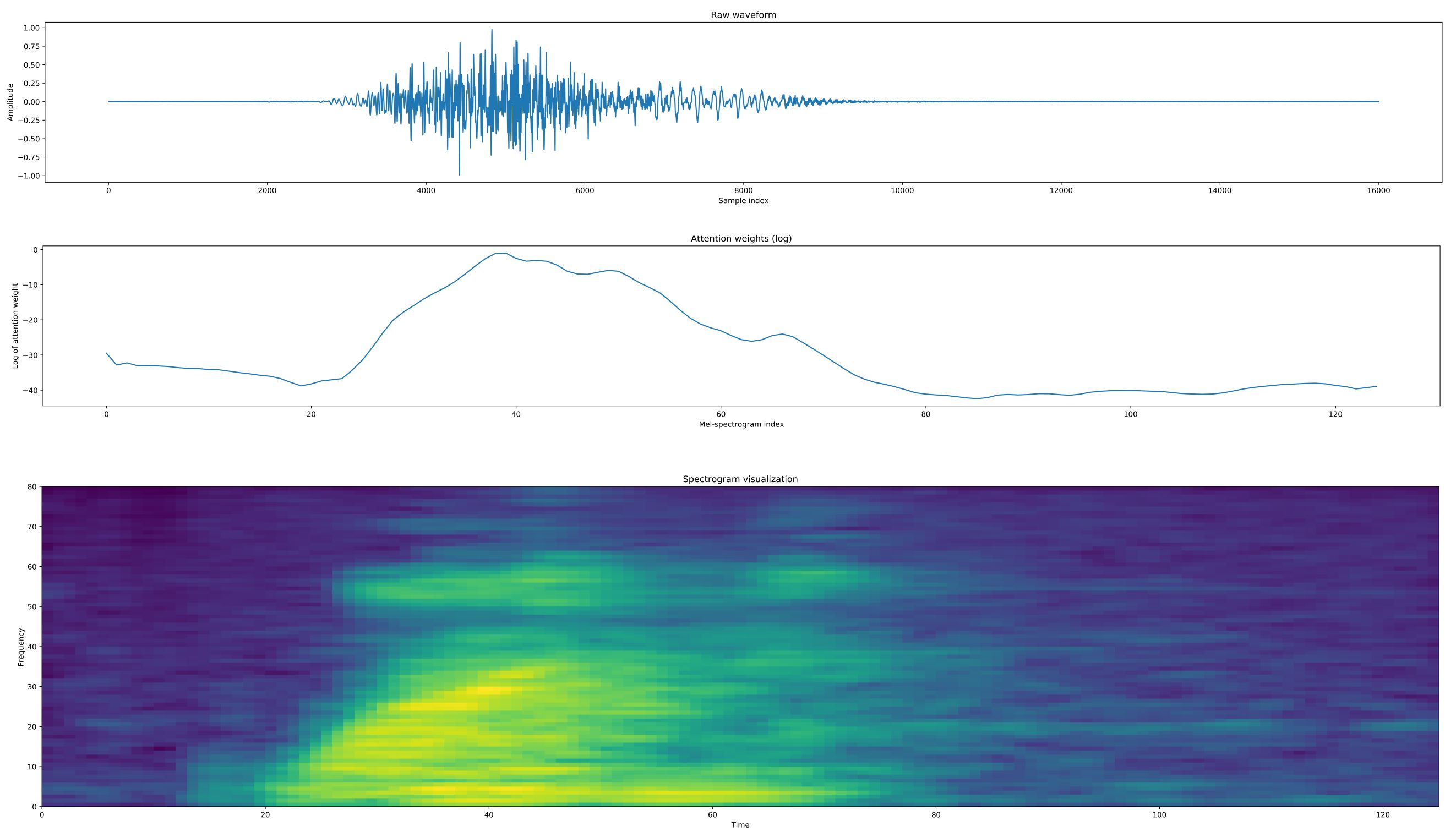}
\caption{Waveform, mel-frequency spectrogram and attention weights for the word ``one''}
\label{figAttONE}
\end{figure}

\begin{figure}[!ht]
\centering
\includegraphics[width=1\textwidth]{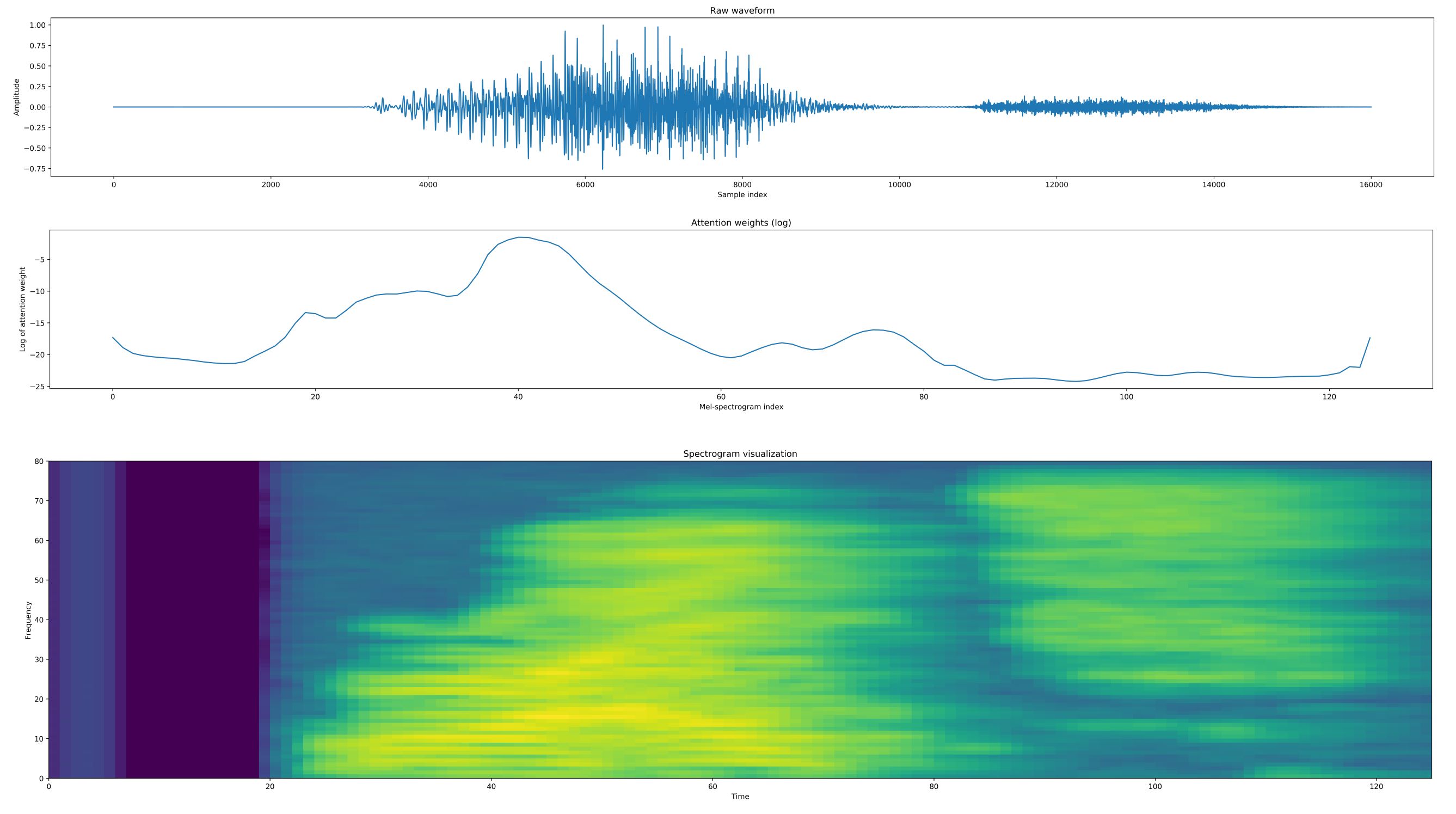}
\caption{Waveform, mel-frequency spectrogram and attention weights for the word ``right''}
\label{figAttRIGHT}
\end{figure}

Figures~\ref{figAttON},~\ref{figAttONE}~and~\ref{figAttRIGHT} show attention weights along with waveform and mel-scale spectrogram. For better visualization, attention weights were plotted in a log-scale. Intuitively, one would expect that the network should put more emphasis on vowel transitions, since non-voiced regions (consonants) may be confused with background noise or even not present at all and regions where vowels are sustained do not carry extra information, i.e., a speaker might pronounce the /a/ sound in right for a long time before transitioning to the /i/ sound. Note that the model matches this intuition and attributes the highest weight to transitions.

\subsection{Confusion Matrices}
\label{secConfMatrices}

\begin{figure}[!ht]
\centering
\includegraphics[width=1\textwidth]{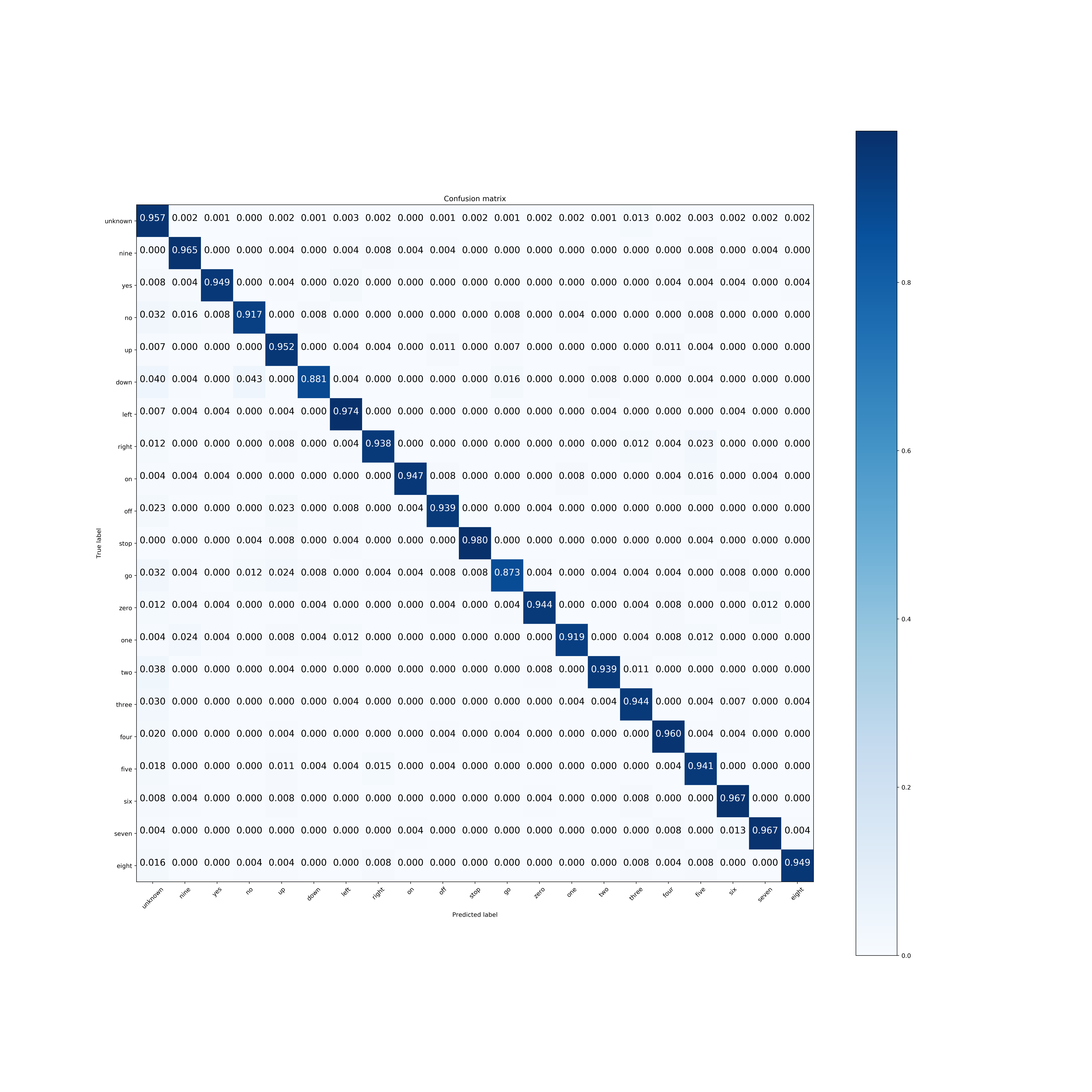}
\caption{Confusion matrix for the 20 command task on Google Speech Dataset V1}
\label{figConfMatrix20cmdV1}
\end{figure}

\begin{figure}[!ht]
\centering
\includegraphics[width=1\textwidth]{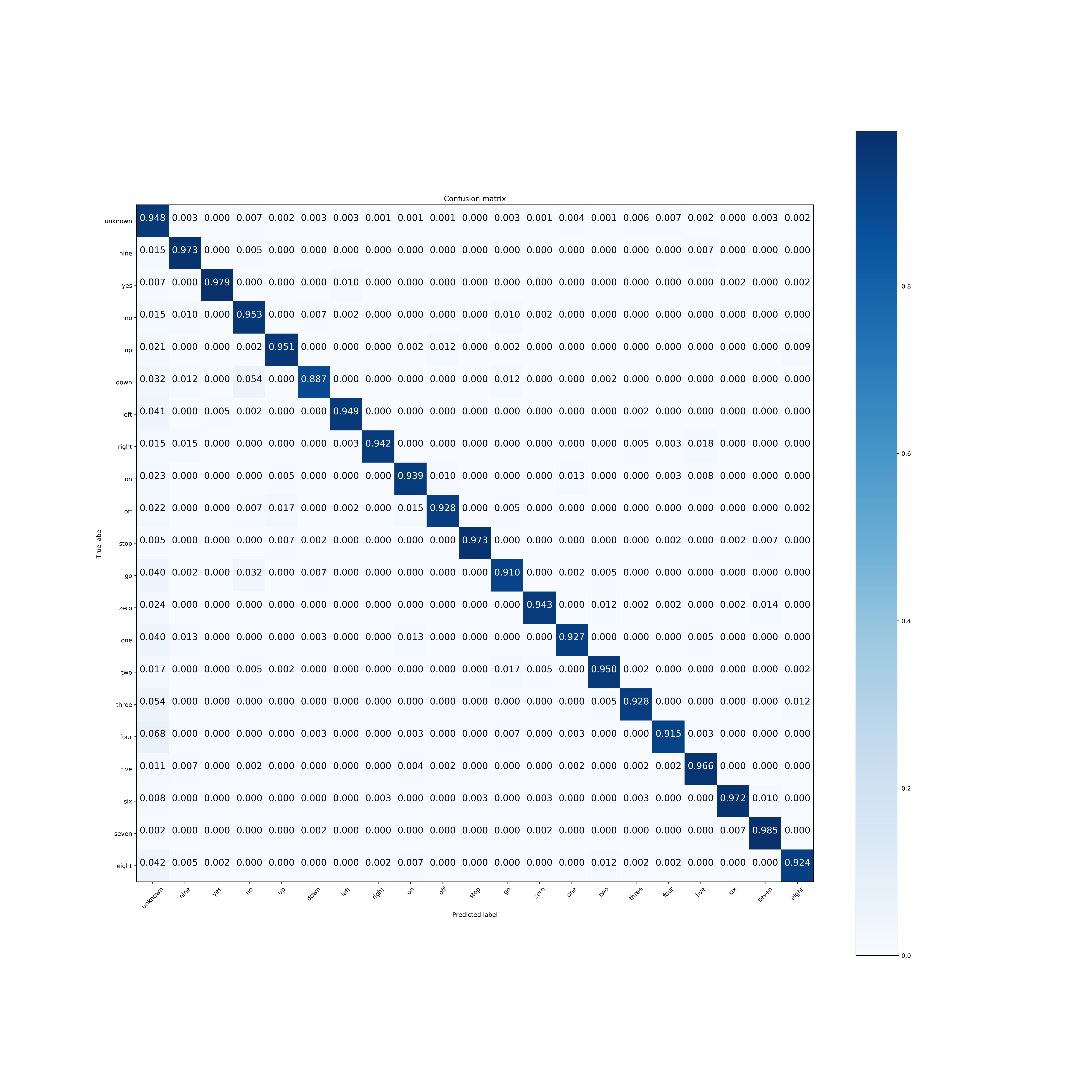}
\caption{Confusion matrix for the 20 command task on Google Speech Dataset V2}
\label{figConfMatrix20cmdV2}
\end{figure}

\begin{figure}[!ht]
\centering
\includegraphics[width=1\textwidth]{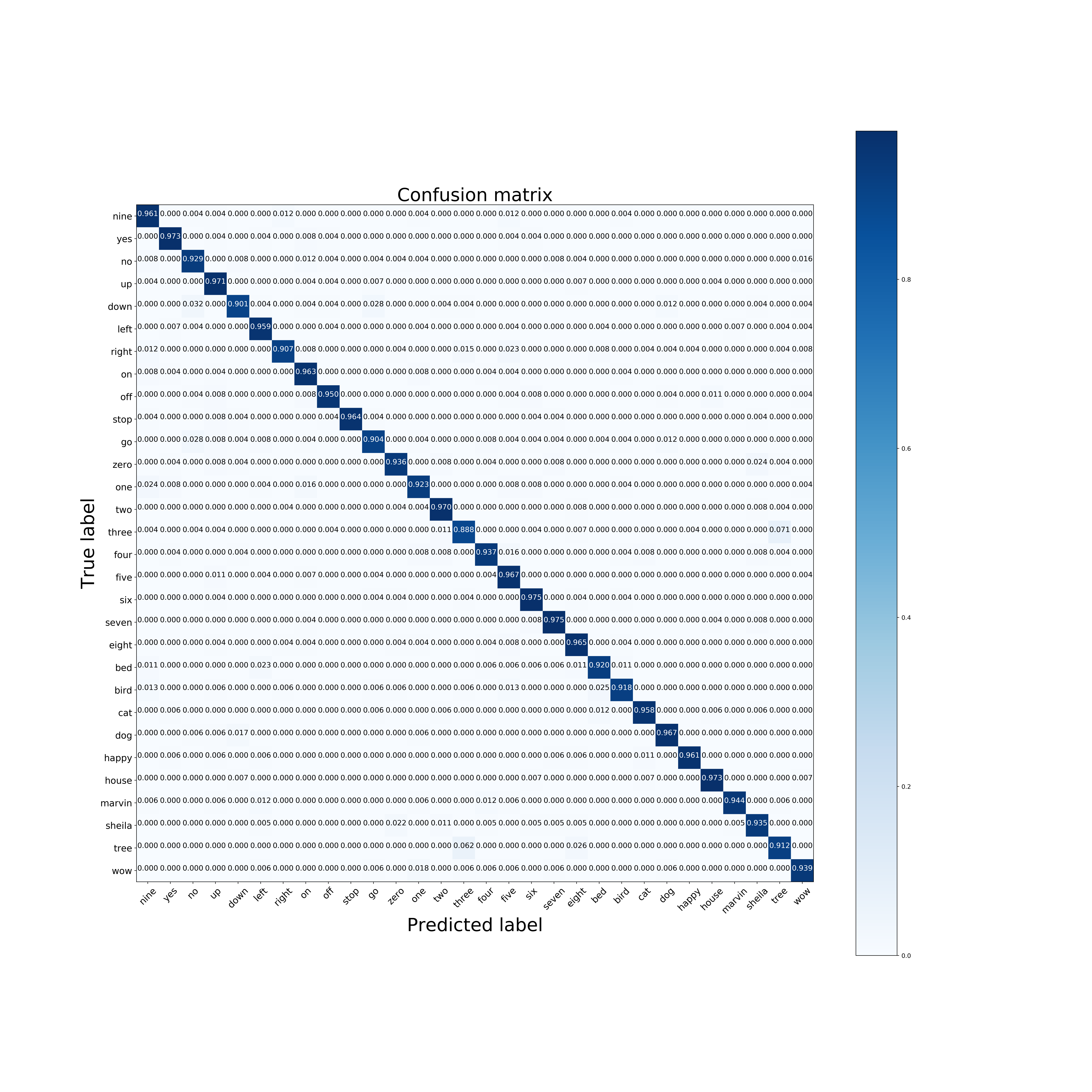}
\caption{Confusion matrix for the 35 word task on Google Speech Dataset V1}
\label{figConfMatrix35V1}
\end{figure}

\begin{figure}[!ht]
\centering
\includegraphics[width=1\textwidth]{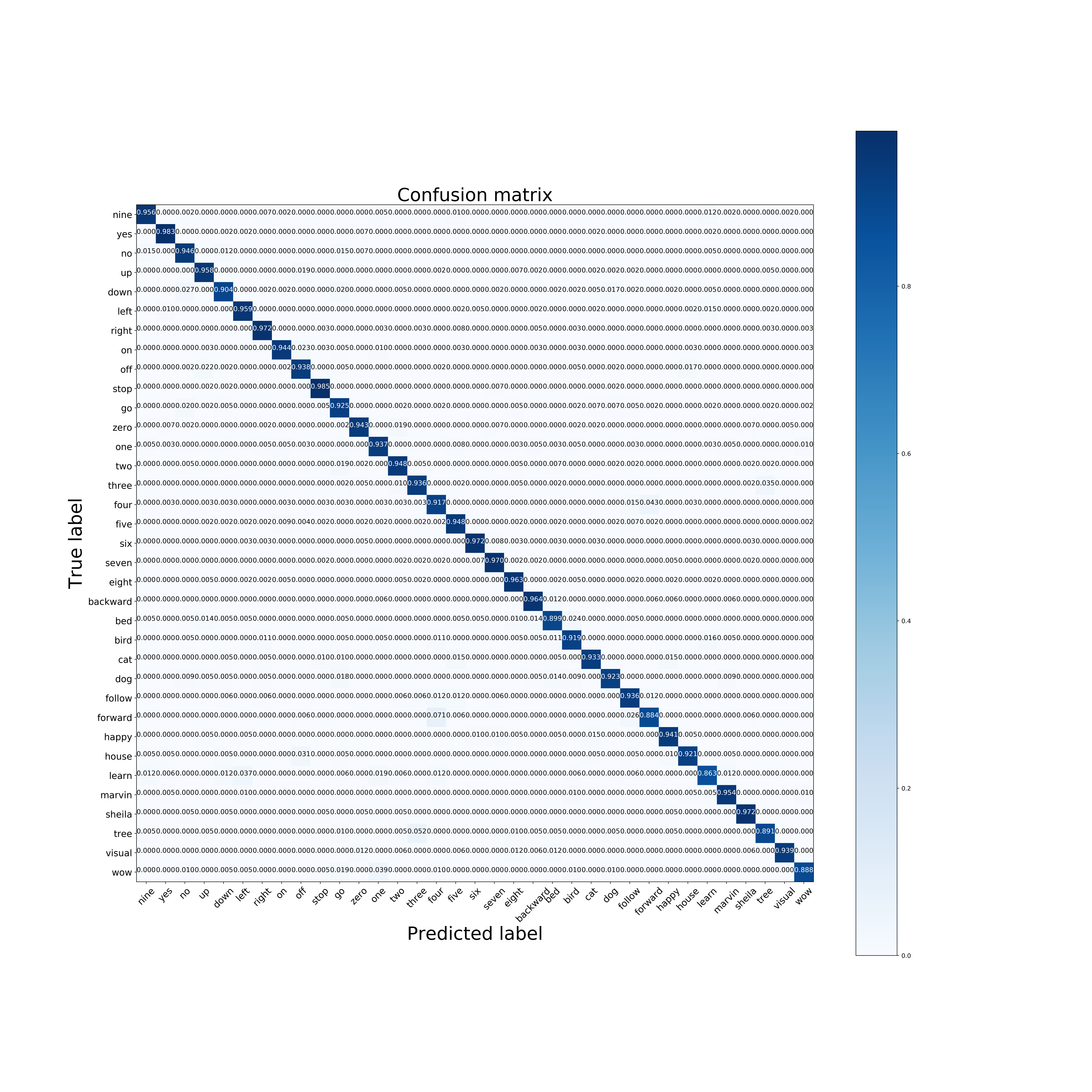}
\caption{Confusion matrix for the 35 word task on Google Speech Dataset V2}
\label{figConfMatrix35V2}
\end{figure}

Confusion matrices are presented for the 20-cmd task and the 35-word recognition task on Google Speech Command dataset V1 and V2 (Figures~\ref{figConfMatrix20cmdV1},~\ref{figConfMatrix20cmdV2},~\ref{figConfMatrix35V1}~and~\ref{figConfMatrix35V2} respectively). It is worth noting that words ``three'' and ``tree'' are often confused (which is expected given the similarity of the words), as well as ``no'' and ``down''. Proper identification of these words would require contextual information from other words in a sentence. This issue should not be relevant for an engineering application where the designers are allowed to pick possible commands: the choice of ``three'' and ``tree'' as possible commands would certainly be a poor choice due to how similar the words are (and also to the fact that non-native speakers sometimes are not even able to pronounce the $/\theta/$ sound in \textbf{th}ree). The accuracy on those words is the lowest (approximately 90\%). Note that in the 35-word task the accuracy on Speech Dataset V2 is $93.9\pm0.2\%$, slightly lower than V1 ($94.3\pm0.2\%$), because V2 has 35 words in the test set (as opposed to 30 in V1).

\clearpage

\section{Conclusion}
\label{secConclusion}

Speech command recognition is present in a wide range of devices and utilized by many HCI interfaces. In many situations, it is desirable to obtain high accuracy, lightweight models that can run locally. In this work, we introduced a novel attention RNN architecture that achieves state-of-the-art performance on multiple KWS tasks while still keeping a small footprint in terms of trainable parameters. Source code is made available on github (to be posted) to enable further work.

The proposed architecture uses raw WAV files as inputs, computes mel-scale spectrogram using a non-trainable Keras layer, extracts short and long-term dependencies and uses an attention mechanism to pinpoint which region has the most useful information, that is then fed to a sequence of dense layers.

The Google Speech Commands datasets V1 and V2 \cite{DBLP:journals/corr/abs-1804-03209} are used to demonstrate the effectiveness of the attention RNN approach. Attention RNN establishes a new state-of-the-art result on all tasks: 20-cmd, 12-cmd, 35-word and left-right. The accuracies are respectively 94.1\%, 95.6\%, 93.9\% and 99.2\% on the V1 dataset and 94.5\%, 96.9\%, 93.9\% and 99.4\% on the V2 dataset.

In engineering applications, being able to explain \emph{what features} were used to select a particular category is a desirable element that is not available in previous neural network models. Our results demonstrate that the attention mechanism explains what parts of the audio are important for classification and also matches the intuition that regions of vowel transitions are relevant to recognize words. For completeness, confusion matrices are presented and show that the word pairs tree-three and no-down are difficult to identify without extra context.

Although data augmentation has been proven to be an important tool to increase model accuracy in visual tasks, the effectiveness of augmenting the audio samples with noise from other datasets was not explored. One possible direction of future work is to investigate the effect of incorporating multiple datasets and using pretrained models. It should also be possible to stack pairs of words for more complex commands and use a sequence-to-sequence model or multiple attention layers. Further investigation will be conducted towards using the proposed attention RNN model for automatic language identification and detection of speech pathologies from audio.



\bibliographystyle{elsarticle-harv}

\bibliography{ABibliography}

\appendix


\end{document}